\documentstyle[aps]{revtex}

\begin{document}

\title{Magneto-optical evidence of the percolation nature of the
metal-insulator transition in the 2D electron system}

\author{I.V. Kukushkin$^{a,c}$, Vladimir 
I. Fal'ko$^{b,c,a}$, R.J. Haug$^{a,d}$, K. v.
Klitzing$^a$, and K. Eberl$^a$}

\address{ $^a$ Max-Planck-Institut f\"ur Festk\"orperforschung, 
Heisenbergstr. 1, 70569 Stuttgart, Germany \\
$^b$ School of Physics and Chemistry, Lancaster University, 
LA1 4YB, Lancaster, UK \\
$^c$ Institute of Solid State Physics, RAS, Chernogolovka, 142432 
Russia \\
$^d$ Inst. f\"ur Festk\"orperphysik, University of Hannover, 
Appelsrt. 2, 30167 Hannover, Germany}

\maketitle

\begin{abstract}

We compare the results of the transport and time-resolved 
magneto-luminescence measurements in disordered 2D electron 
systems in GaAs-AlGaAs heterostructures in the extreme quantum limit, 
in particular, in the vicinity of the metal-insulator transition (MIT).  
At filling factors $\nu <1$, the optical signal has two
components: the single-rate exponentially decaying part attributed
to a uniform liquid and a power-law long-living tail specific to a
microscopically inhomogeneous state of electrons.  We interprete this
result as a separation of the 2D electron system into a liquid and
localized phases, especially because the MIT occurs strikingly close
to those filling factors where the liquid occupies ${1\over 2}$ of the
sample area (the percollation threshold condition in two-component
media).  

\end{abstract}

\narrowtext

\twocolumn

The metal-insulator transition (MIT) in two-dimensional (2D) electron 
systems still remains to be a hot topic in the semiconductor physics 
\cite{MIT1,Glozman,MIT2}, in particular, because of a variety 
of transitions between quantum Hall liquids and insulating states in 
the extreme quantum limit realized at high magnetic fields. 
Depending on the level of disorder in the 
system, the MIT can be either due to the Wigner cristallization, or 
due to localization.  In pure enough systems, it has the thermodynamic 
character and is conventionally attributed to the interaction-induced 
crystallization of 2D electrons into the Wigner solid
\cite{MIT2}.  In disordered structures, the dielectric states is 
formed because of localization of the 2D electrons by a random 
potential coming from an inhomogenuous charge distribution in the 
donor layer \cite{MIT1,Glozman}, and the interest of the present
publication is focused on the latter situation.
 
Although the single-particle scenario of localization of electrons at 
high magnetic fields was already thoroughly explored some years 
ago \cite{Pruisken}, the most recent theories \cite{MIT4} emphasize
the influence of interactions on the MIT also in disordered systems,
especially regarding the existence of a hierarchy of strongly
correlated quantum Hall liquids.  Being inspired by the success of
the scaling ideas developed in single-particle models \cite{Pruisken}, 
the search of   
a global phase diagram of the MIT in the extreme quantum limit has 
been started \cite{MIT4} in order to describe the growing amount 
of experimental data.  However, one should not ignore such a 
possibility that, in interacting electron systems, the transiton 
between correlated liquids and disordered insulator is the Mott 
transition rather than the Anderson one, and that the mesoscopic-scale 
inhomogenuity in the strength of a random potential itself can result 
in the separation of the 2D system into phases with 
locally different properties 
(liquid versus localized insulator).  In this scenario, the portion of 
a localized phase varies contineously under the variation of a magnetic 
field due to the sensitivity of the electron-electron correlations to 
the value of the filling factor, so that the {\it MIT detected in the 
transport} might manifest nothing but the {\it destruction of the 
percolation through the liquid} in a multicomponent media \cite{Perc}.  

To find the truth, it is necessary to make a local measurement which 
would distinguish between different phases (liquid and insulator) and
to evaluate amount of each of them.   
We solve this problem by using the method based on the
analysis of the time-resolved recombination of 2D electrons with a
small number of acceptor bound holes in the $\delta$-doped layer.  
We already know \cite{KukuFalkoPRL} that, in contrast
to the quantum Hall effect state at the filling factor $\nu =1$ that 
demonstrates the single-exponential decay $I(t)\propto\exp (-t/t_0)$,
the magnetically frozen state manifests itself in a broad spectrum of
recombination rates which results in the power-law kinetics of the
magneto-luminescence.  It is a peculiarity of the recombination
process involving acceptor bound holes that the latter form a small
neutral centers, and that the life-time $\tau$ of each hole is
determined by the local density of 2D electrons at the acceptor
position, $\tau^{-1}\propto\rho_e$, so that the spectrum
$f(\tau^{-1})$ of recombination rates $\tau^{-1}$ reflects the
distribution of the local 2D electron density \cite{F}.  The latter
can be found from the inverse Laplace transform of the measured
temporal dependence $I(t)$ of the luminescence intensity after a short
illumination pulse, 
$I(t)\propto\int d\tau^{-1}f(\tau^{-1})e^{-t/\tau}$.  
For example, the mentioned above single-exponential decay of the
luminescence at $\nu =1$ evidences the microscopic-scale homogeneity
of the 2D electron density in this incompressible liquid state
\cite{Janovici}, whereas the information about the distribution of
local values $\rho_e$ which we extract from the long-living tails of
luminescence from any state at $\nu <1$ may be used for identification
of its intrinsic structure, like it has been done before in the Wigner
solid regime \cite{KukuFalkoPRL}.

In the present paper we combine the transport measurements with
the time-resolved optics \cite{KukuFalkoPRL,KukuFalkoJETP}
to study systematically a series of heterostructures with various
densities and level of disorder.  In the entire interval of filling
factors $\nu <1$ and in all the samples, the magneto-luminescence
kinetics $I(t)$ looks as if it is composed of a single-exponential
(liquid-type) contribution $I_l(t)\propto\exp (-t/t_0)$ and the
universal contribution $I_i(t)$ from inhomogeneous localized state
with a power-law long-living asymptotics $I_i(t)\propto
(t_0/t)^{\alpha}$ summed with different weights.  This 
enables us to distinguish between the liquid and localized phases and
to estimate their portions in the mixture.  Our result shows that for
all the samples under investigation the liquid phase fills
about ${1\over 2}$ of the sample area at the MIT point, which coincides
with the condition of the percolation threshold in 2D
two-component media \cite{Perc}.  

We studied six different quality (see Table 1) GaAs-AlGaAs single 
heterojunctions with a $\delta$-doped monolayer of Be acceptors 
($n_A=1.3\times 10^9cm^{-2}$) located in the wide ($1\mu m$) GaAs
buffer layer at a distance about 30 nm from the interface
\cite{Old-K}. The magnetoresistance tensor was measured on the Hall
bar samples by use of a standard low-frequency (12 Hz) and low-current
(10 nA) lock-in technique with a resistance $100M\Omega$ connected in
series with the sample.  For photoexcitation, we used pulses from a
tunable Ti-Sapphire laser (wavelength $\sim$ 800nm) with duration 20
ns, peak power of $10^{-5}$ $-10^{-4}$ $W/cm^2$ and frequency of
$10^5$-$10^3$Hz \cite{KukuFalkoPRL,Old-K}.  

Fig.  1 shows the results of magneto-transport measurements (in the 
range of temperatures $T=50-150mK$) which were used for  
determining the MIT point in two samples B and E which have 
identical concentration but very different mobilities (see Table 1).  
The conductivity measured in samples of different qualities shows 
quite distinguishable behavior, especially at $\nu <1$.  In particular, in 
better quality samples (A,B,C) one can find the inter-crossing between 
$\sigma_{xx}(B)$ taken at different temperatures which can be used for 
determination of the MIT point \cite{MIT1}.  However, such an 
inter-crossing does not appear in the magneto-conductivity data taken 
in the lowest mobility structures (D,E,F), so that another criterion 
of the MIT is desirable.  The common feature of all studied samples
near the MIT was an abrupt appearance of the out-of-phase signal both
in the diagonal and Hall components of resistivity which is
accompanied by a drastic decrease of a current and indicates the
formation of a dielectric state.  Fig.  1 also 
shows the evolution of the magnetic field dependence of the current
through  the 2D system with the temperature.  Under the decrease of a
filling factor, the current shows a threshold at $\nu_c$ which becomes
more pronounced at lower temperatures, and treate it 
below as the indication of the  MIT point.  The value of $\nu_c$
determined in such a way is pretty close to the position of the peak
in $\sigma_{xx}$ in samples D-F, which has been used by Glozman {\it
et al} and others \cite{Glozman} as a criterium of MIT.  Note that in
the present investigation we do not fight for the accuracy of
determining $\nu_c$ better than few percent, since our aim to see a
correspondence between the MIT data observed transport and optical
data that show no threshold behavior.  Indeed, in contrast to the
transport, the optical data taken in the same samples demonstrate
only a continuous evolution both of the integral and time-resolved
spectra.  Insets to Fig.  1 (a) and (b) show the raw data of typical 
recombination kinetics measured in samples (B,E) in different 
magnetic fields at $T=50mK$ and we see no critical changes 
occuring when $\nu =\nu_c$ is passed.  

The magneto-luminescence kinetics measured in lower- and
higher-mobility samples has both common features and significant
differences.  The common feature of the recombination kinetics
observed in all the samples is the purely exponential decay,
$I(t)\propto\exp (-t/t_0)$, of the magneto-luminescence intensity at
the filling 
factor $\nu =1$.  Since the density of optically active holes is much
smaller than the 2D electron density, the time dependence $I(t)$ is
controlled by the recombination rates of holes.  In the homogeneous
electron system, the latter are determined by the sheet electron
density and the distance from the interface to the acceptor
mono-layer and are the same for all of them, so that the single-rate 
($\tau^{-1}=t_0^{-1}$ exponential decay manifests the microscopic scale
homogeneity of the electron density.  At $\nu =1$, this is the well
known property of a trial many-body wave-function corresponding to a
completely filled Landau level \cite{Janovici}.

For the lower fillings, the slow recombination tail develops, which 
indicates the formation of a short-range scale inhomogenuity of the 
local electron density (on the scale of $n_e^{-1/2}$).  This behavior
is common for all samples we studied and evidences a partial 
localization of the system under an increase of a magnetic field 
which starts much earlier than the MIT takes place.  Here, 
the difference between low and high mobility samples manifests 
itself:  in the former, the long-living tails develop at considerably 
higher filling factors (lower magnetic fields).  To make a comparison 
easier, in Fig.  2 we show the kinetics of magneto-luminescence 
$I(t)$ for three groups of samples 
($\{B,D\},\{B,C,E\},\{E,F\}$), each group been studied at the same 
filling factor ($\nu =0.19;0.42;0.58$).  Within each group of 
samples, the luminescence intensity presented in Fig. 2 is normalized 
to the same integral value (e.g.,  
$\int I^{(B)}(t)dt=\int I^{(D)}(t)dt$ in the first group), which
enables us to compare both the form of the long-living tails 
and their strength in samples with different parameter and 
at the same or different values of the filling factor.   

First of all, after comparing the data taken in different samples at 
the same $\nu$, to say, $B,C$ and $E$ for $\nu =0.42$, we find that 
the strength of the long-living recombination tail is systematically 
more significant in lower quality samples.  At the same time, the form
of the tails at a long enough delay time is essentially the same for 
different samples studied at the same values of the filling factor. 
The presentation of the data in Fig.  2 using the log-log plot has been
chosen in order to demonstrate that the recombination kinetics at the 
long-living tails obey the power law \cite{KukuFalkoPRL,KukuFalkoJETP}, 

\begin{equation}\label{e1}
I(t)=A\left(t_0/t\right)^{\alpha},\quad t>t_0,
\end{equation}
and that the exponents $\alpha$ of the power law dependence coincide 
in different samples, if we make the measurement at the same filling 
factor, although the strength of the tail (i.e., parameter $A$) is 
higher when disorder is stronger.  The power law kinetics has been 
predicted by the earlier theory of recombination in the extreme 
quantum limit \cite{F}, with the exponent 
$\alpha$ which can be approximated by $\alpha\approx 1+\nu$ 
at $\nu\ll 1$.  The fact that the exponent of the magnetoluminescence  
kinetics calculated in Ref. \cite{F} depends only on the value of
the filling factor was the result of the assumption that in the 
magnetically frozen state electrons are strongly localized at the 
interparticle distances - each sitting opposite to its own donor 
behind a narrow spacer.  The spectrum of experimentally derived 
$\alpha$'s for all the available samples and for a broad range
of filling factors, $0.75>\nu >0.15$, is presented in the inset to
Fig.  2.  Although these data correspond to the samples of a very
different quality and concentration, the experimentally obtained
values of $\alpha$ group near a single filling factor dependence 
$\alpha (\nu )$.  This is what we call the {\it universality\/} of
the power-law recombination kinetics of localized electrons.  In a
good approximation, the empirical $\alpha$ has the form $\alpha\approx 
(1-\nu )^{-1}$, which is consistent with the theory by Fal'ko 
\cite{F} for $\nu\ll 1$.

As we discussed above, the kinetics of recombination of 2D electrons 
with acceptor-bound holes in a $\delta$-doped layer is governed by the 
spectrum of the recombination rates of holes, 
$\tau^{-1}\propto\rho_e$, which, in its 
turn, reflects the character of the distribution $f(\rho_e)$ of the 
local 2D 
electron density, $\rho_e$.  From this point of view, the observed 
power-law kinetics of recombination in Eq.  \ref{e1} can be directly 
recalculated (using the inverse Laplace transform) into the power-law 
distribution of the local 2D electron density, $f(\rho_e)\propto\rho_
e^{\beta}$ with 
$\beta =\alpha -2\approx (2\nu -1)/(1-\nu )$ and the observed
universality of the  
luminescence kinetics has behind it the universality of the 
distribution of a local density in a localized disordered state of 
magnetically frozen electrons.  Below, we use this fact for measuring 
the disorder-dependent portions of insulating and liquid phases in the 
sample.  Fig.  3 shows that the extraction of the power-law tail 
from the total intensity gives the rest of it which has a typical 
single-exponential behavior at the intermediate range of the delay 
time. The parameters ($A$ and $\alpha$) of the slow recombination tail 
$I_i(t)=A(t_0/t)^{\alpha}$ were determined from a fit of the power law 
asymptotics at $t\gg t_0$.  The rest of the recombination intensity, 
$I_l=I(t)-I_i(t)$, is represented in Fig. 3 by solid circles and 
demonstrates an exponential decay, $I_l(t)\propto\exp (-t/t_0)$ 
which can be followed over the same number of decades as at $\nu =1$.  
The single-rate decay is the property of a microscopically 
homogeneous density, so that we assign this part of the luminescence 
intensity to the liquid state of electrons and suggest that the 2D 
electron system represents a mixture of a liquid and disorder-induced 
localized phases.  

The goal of our analysis is to construct a diagram 
illustrating the evolution of the portion of the sample area, 
$S_l(\nu )=\left[\int I_ldt\right]/\left[\int Idt\right]$, 
occupied by a liquid under the decrease of 
the filling factor.  The dependence $S_l(\nu )$ determined for six  
different samples is shown on Fig.  4.  The relative amount of the 
liquid phase in all the structures tends to decrease under an increase 
of a magnetic field, which manifests the formation of the freeze-out 
regime.  In the low-mobility samples (D,E,F), the localization is much
more efficient and develops smoothly in the entire region of filling
factors $\nu <1$.  

In the better quality samples (A,B,C), the general tendency of the 
portion of the liquid to decrease with increasing of a magnetic field 
is accompanied by oscillations associated with the fractional 
quantum Hall effect states (see Fig.  4).  The amount of the 
recombination events that we assign to the liquid has clear maxima at 
$\nu =\frac13$ in samples B and C, and at $\frac13,\frac15,\frac25$ in 
the sample A.  This is a manifestation of the incompressibility 
of the Laughlin liquid, in which the interaction effects are strong 
enough (compared to disorder) to keep the electron density 
microscopically homogeneous \cite{Janovici}.  This supplies us 
with an additional argument in favor to the 
assumption that the separation of the system into the liquid and 
insulating phases at any filling factor is the result of a competition 
between electron-electron interactions and disorder in which the 
former tempt to form a homogeneous liquid state whereas the
latter tends to break correlations among electrons and to localize
them.  Note, that even the best quality samples studied in the present
paper were disordered enough and did not show abrupt features specific
to the Wigner crystallization \cite{KukuFalkoPRL}.   

Let us finally focus at the region near the critical filling factor
$\nu_c$ of the MIT determined using the transport measurements (see
Fig. 1).  For each single sample, the optically characterized level of 
localization smoothly changes near $\nu_c$ and shows no striking
features.  However, been compared in a wide variety of structures, the
MIT always occurs strikingly close to those filling factors where the 
continuously varying portion of a liquid fills one half of the sample
area.  The fundamental property of the percolation transition in
macroscopically inhomogeneous 2D media is that it occurs when the
insulating component occupies one half of the sample area, that's when
the percolation path through the conducting phase is destroyed
\cite{Perc}.  Our conclusion is that the magnetic field driven
MIT which we observe at $\nu <1$ is related to the {\it percolation
threshold} through the liquid in the systems with the phase
separation into a liquid and disorder insulator.  

{\it Acknowledgement.}  The authors thank Volkswagen Stiftung and
Russian Fund of Fundamental Research for a financial support.  One
of us (VF) also acknowledges a support from EPSRC and NATO CRG 921333.

\begin{figure}
\caption{Magnetic field dependence of the conductivity and current 
measured in samples B (a) and E (b) for different temperatures ((a) - 
$T=150;\,100;\,50mK$, and (b) - $T=100;\,50mK$).   The critical
filling factor $\nu_c$ of the MIT is indicated.  Insets represent the
raw data of the time-dependent magneto-luminescence kinetics measured in
the same samples at filling factors $\nu <1$ at $T=50mK$. \label{Fig1} }
\caption{ \label{Fig2} Comparison of the recombination kinetics in 
groups of samples taken at three different filling factors and at 
$T=50mK$.  The integral luminescence intensity is normalized by 
the same value within each group.  Dashed lines illustrate a 
draft fit of the long-living tail using a power law.  Inset shows the
dependence of $\alpha^{-1}$ on filling factor measured in all the  
studied samples. }
\caption{ \label{Fig3} Example of a separation of the time-dependent
magneto-luminescence intensity $I(t)$ into the power-law and 
exponential parts, $I=I_i+I_l$. Solid line correspond to 
the fit of the long-living tail of $I(t)$ by the power law 
$I_i(t)=A(t_0/t)^{\alpha (\nu )}$.  Solid circles and dashed line
are the dependence $I_l(t)=I(t)-I_i(t)$ at an interemediate
range of the delay time and a single-exponential fit to it.  } 
\caption{ \label{Fig5} The filling factor dependence of a portion of
the sample area occupied by a liquid in all studied structures.   
Black squares indicate the points taken exactly at the filling
factors of the MIT detected in the transport measurements.
Dashed line and arrows illustrate the classical condition of a
percolation threshold in a two-component systems mixed of a 
metallic and insulating phases.}

\end{figure}


\begin{table}

\caption{Samples parameters: density $n_e$, mobility $\mu$,
$t_0$ \label{Table1}} 

\begin{tabular}{ccccccc}     
sample &A &B &C &D &E &F                            \\
$n_e$ ($\times 10^{11}cm^{-2}$) 
              &0.55 &0.60 &0.80 &0.71 &0.60  &1.43  \\
$\mu$ ($\times 10^5cm^2/Vs$) 
              &12.1 &3.85 &0.94 &0.43 &0.21  &0.11  \\
$t_0$ ($\times 10^{-7}s$)
              &1.9  &1.9  &2.3  &2.2  &2.0   &2.9   \\
\end{tabular}

\end{table}

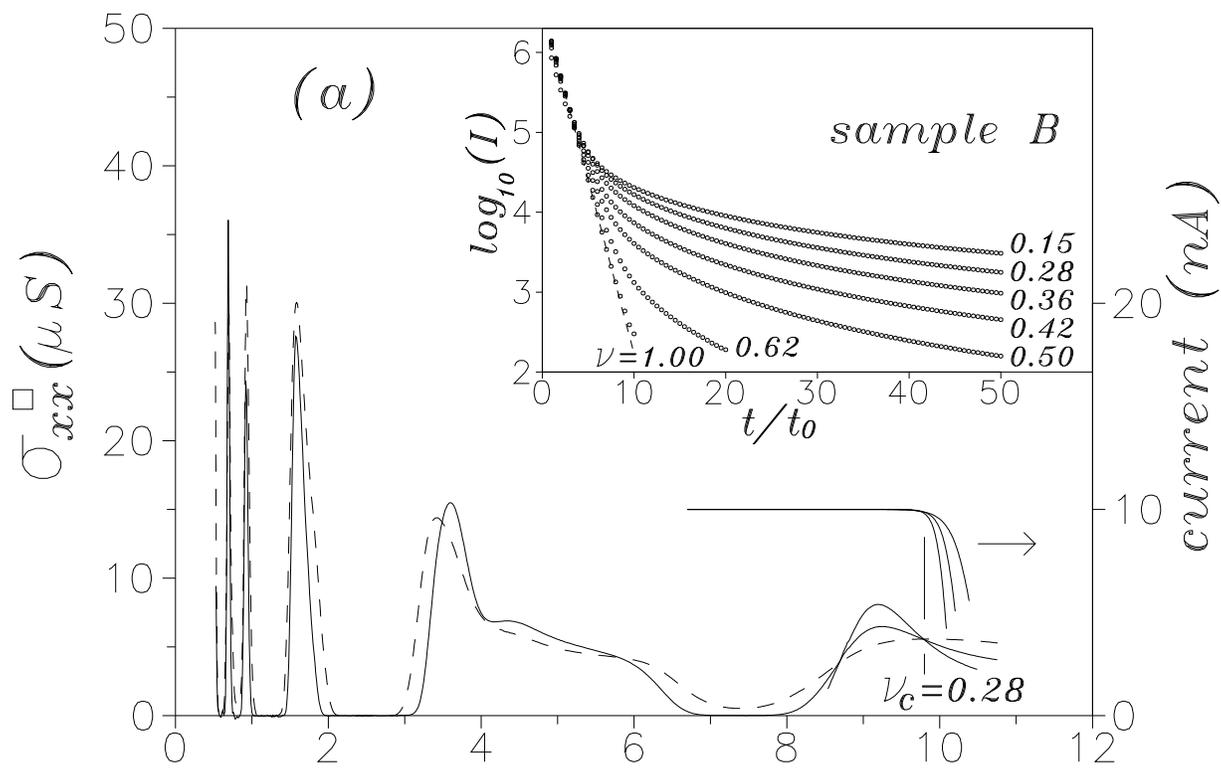
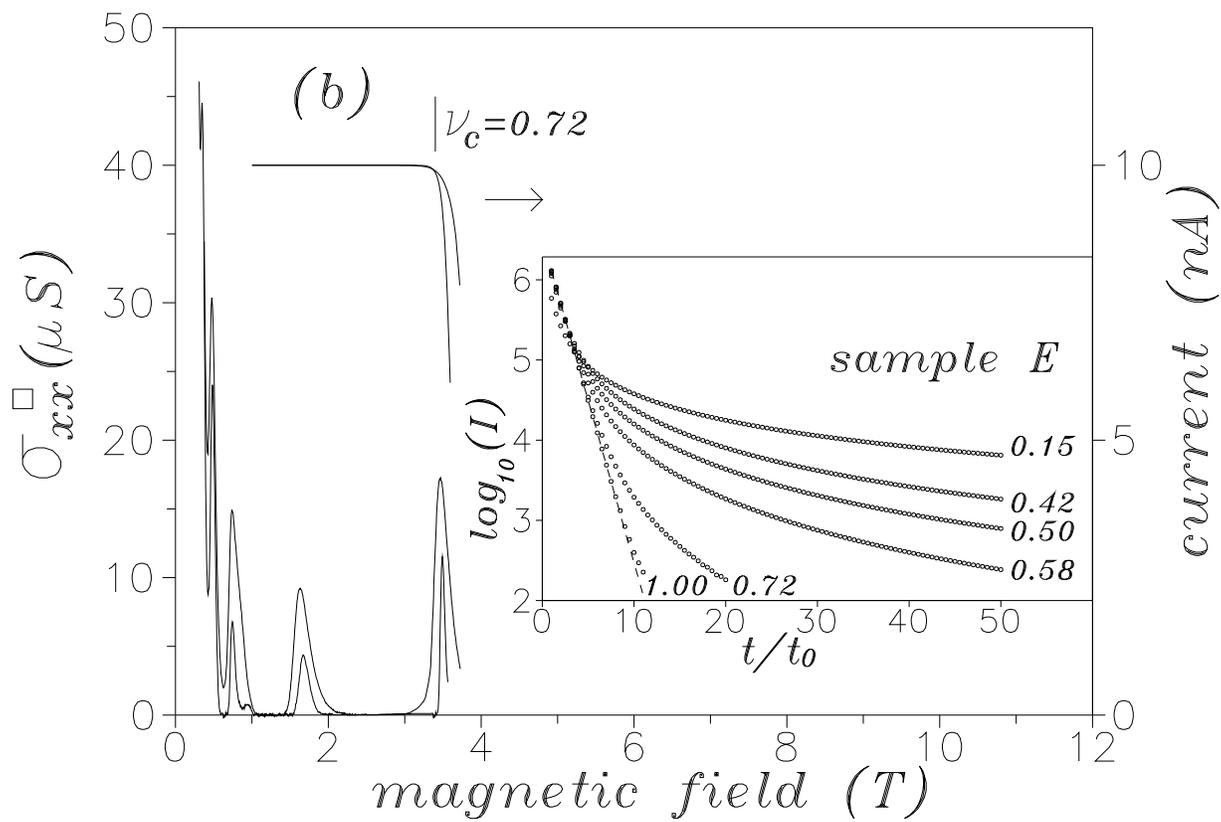
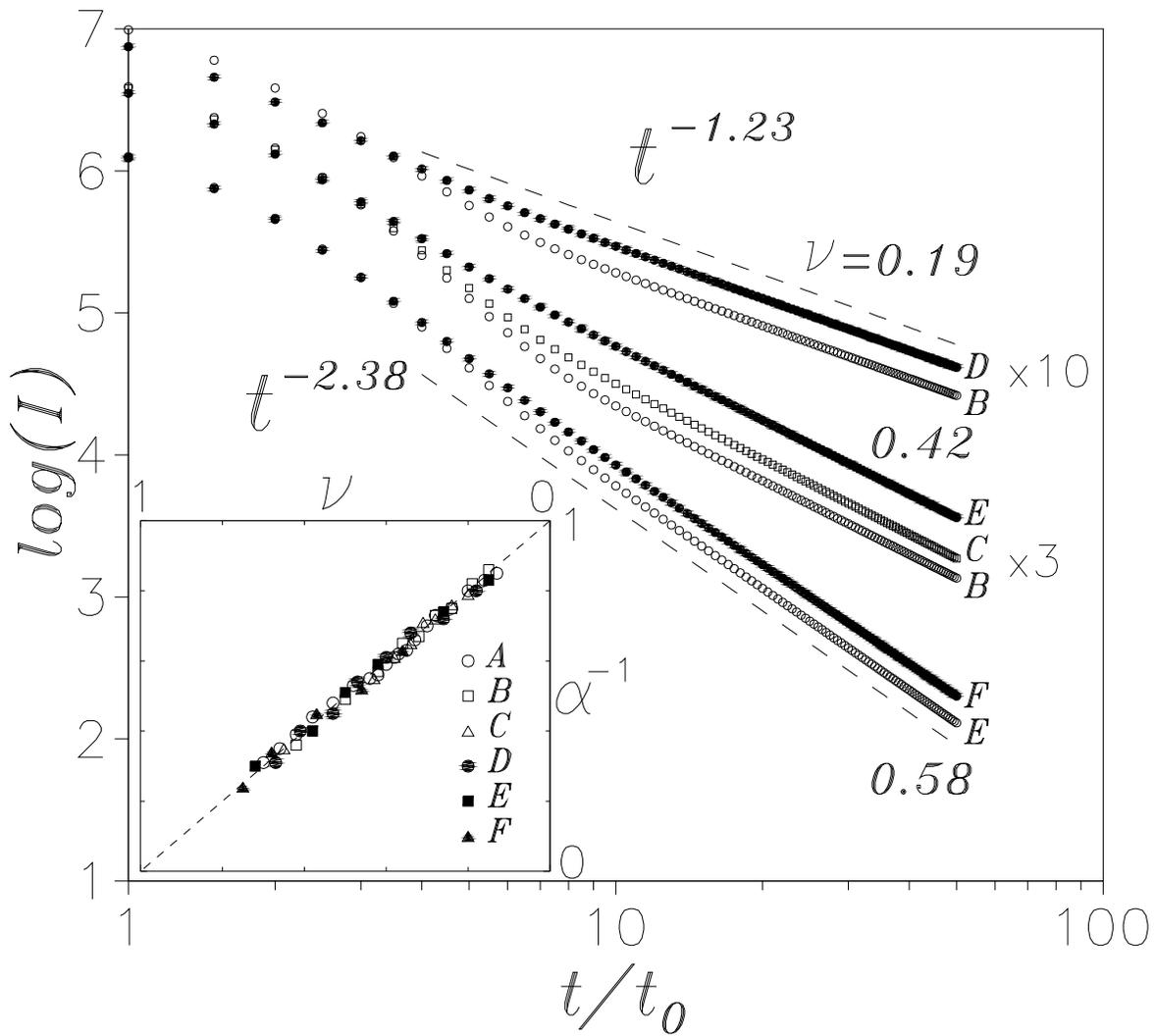
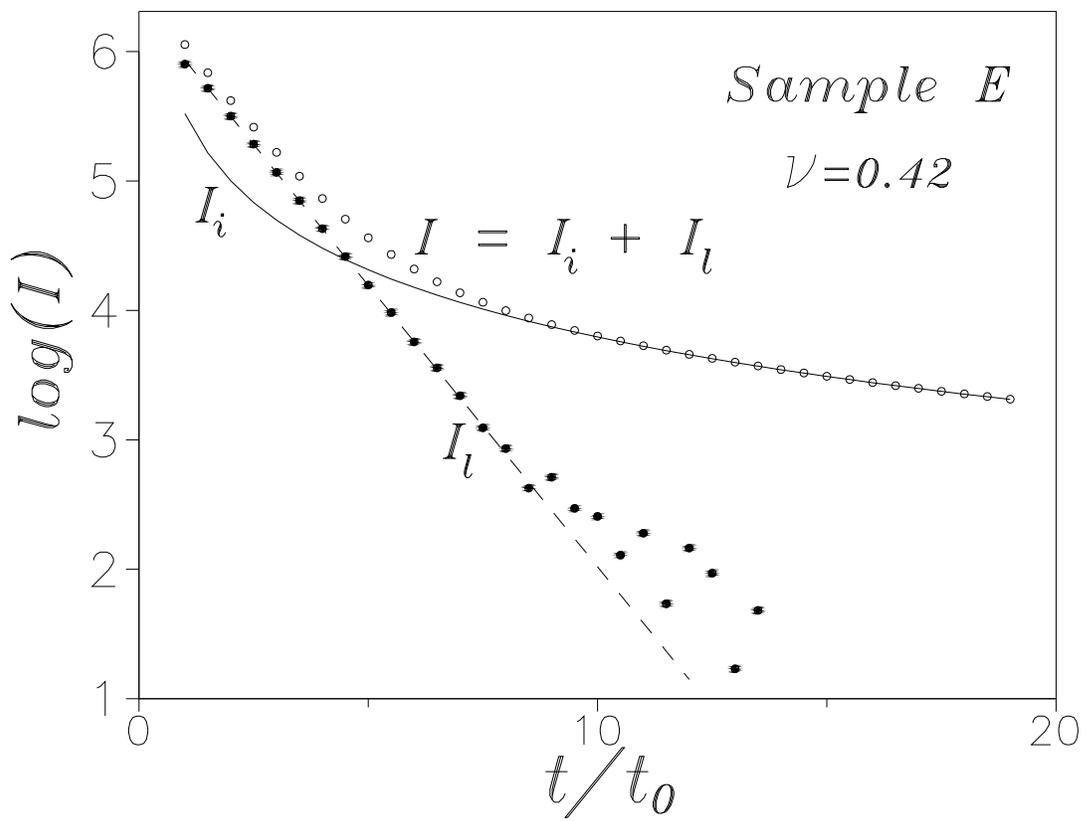
\end{document}